\documentclass[letter]{ieice}
\usepackage[dvips]{graphicx}
\usepackage[varg]{txfonts}

\field{B}
\title{SNR Degradation due to Carrier Frequency Offset in {OFDM} based Amplify-and-Forward Relay Systems}
\authorlist{% fill arguments of \authorentry, otherwise error will be caused.
 \authorentry[ yxjiang@seu.edu.cn]{ Yanxiang~Jiang}{m}{labelA}[labelB]
 \authorentry{ Yanxin~Hu}{n}{labelA}
 \authorentry{ Xiaohu~You}{n}{labelA}
}
\affiliate[labelA]{The authors are with the National Mobile Communications Research Laboratory,
Southeast University, Nanjing 210096, China. }
\paffiliate[labelB]{Presently, the author is with the }

\received{2010}{10}{21}
\begin{document}
\maketitle
\begin{summary}
In this letter, signal-to-noise ratio (SNR) performance is analyzed
for orthogonal frequency division multiplexing (OFDM) based amplify-and-forward (AF) relay systems in the presence of
carrier frequency offset (CFO) for fading channels. The SNR expression
is derived under one-relay-node scenario, and is further extended to multiple-relay-node scenario. Analytical
results show that the SNR is quite sensitive to CFO and the sensitivity of the SNR to CFO is mainly determined by the
power of the corresponding link channel and gain factor.
\end{summary}

\section{Introduction}

     Orthogonal frequency division multiplexing (OFDM) technology
is receiving increasing attention in recent years due to its
robustness to frequency-selective fading and its subcarrier-wise
adaptability \cite{Bingham}. It is likely that OFDM will
become a key element in the future wireless communication systems.
On the other hand, wireless relay technology is
highly envisaged due to its capability to support high data rate
coverage over large areas with low costs \cite{Pabst}. In general, there
are mainly two modes of relay techniques, i.e., amplify-and-forward (AF)
and decode-and-forward (DF). In AF mode, the relay only retransmits an amplified
version of the received signals. This leads to low computational complexity
and low power consumption for relay transceivers.
Accordingly, OFDM based AF relay systems have gained much interest in wireless communication
research area \cite{Herdin,Ho,Kaneko}.

    It is well known that conventional point to point OFDM systems are highly sensitive to
CFO. The effect of CFO on point to point OFDM systems has been
investigated in \cite{Pollet,Nikookar,Hwang,Ma}. In \cite{Pollet},
the signal-to-noise ratio (SNR) degradation due to CFO was analyzed for additive white Gaussian noise (AWGN) channels. It was
also analyzed for time-invariant multipath channels in \cite{Nikookar} and
shadowed multipath channels in \cite{Hwang}. Moreover, the bit error rate (BER) expression
in the presence of CFO was derived in \cite{Ma}.
While the CFO problem is an extensively studied subject
in point to point OFDM systems, it is still mostly open and much more complicated for research in
cooperative OFDM systems. Due to the reason that the multiple transmissions
from relay nodes are from different locations with different
oscillators, they may have multiple different CFOs that can not be compensated simultaneously
at the destination node. In this letter, we analyze the SNR degradation due to the presence of multiple CFOs
in OFDM based AF relay systems  for general multipath fading channels.
The rest of the letter is
organized as follows. The system model is presented in Section II. In
Section III, the SNR performance of OFDM based AF relay systems is analyzed.
Simulation results are shown in Section IV. Final conclusions are
drawn in Section V.

\section{System Model}

\begin{figure}[!h]
\centerline{\includegraphics[scale=0.45]{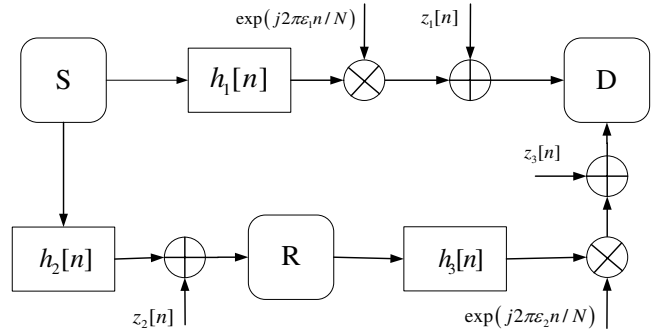}}
\caption{Baseband equivalent system model of an OFDM based AF relay system.}
\label{fig1}
\end{figure}

   We consider the three-terminal model of an OFDM based AF relay system as shown
in Fig. \ref{fig1}, which is composed of a source node S, a destination node D, and a
relay node R. One transmission period is separated into two time slots. In
the first time slot, S transmits an OFDM symbol of N subcarriers to R
and D. In the second time slot, R amplifies and retransmits the OFDM symbol to D, where
the corresponding gain factor is $\rho $.
Further discussion about the gain factor $\rho $ is presented in Section III.

    Let $X\left[ k \right]$ denote the data symbol to be transmitted at source node S.
Let $N_g $ denote the length of the cyclic prefix (CP). Then, with the application of inverse discrete
Fourier transform (IDFT) to $X[k]$ and CP insertion, we have

\begin{eqnarray}
\label{eq1}
\lefteqn{
x[n] = \frac{1}{N}\sum\limits_{k = 0}^{N - 1} {X[k]\exp [j2\pi k{{\left( {n - N_g } \right)} \mathord{\left/
 {\vphantom {{\left( {n - N_g } \right)} N}} \right.
 \kern-\nulldelimiterspace} N}],} \
 }  \qquad\qquad\qquad\qquad \nonumber\\
&&  {\rm{for}} \  0 \le n \le N + N_g  - 1.
\end{eqnarray}

\noindent
In order to prevent possible inter-symbol interference (ISI) between OFDM symbols,
we assume that $N_g \ge L_1$ and $N_g \ge L_2 + L_3 $, where $L_1$, $L_2$, $L_3$ are the
delay spreads of the considered multipath channels $h_1 [n]$, $h_2 [n]$ and $h_3 [n]$ as depicted in
Fig. \ref{fig1}.

    At destination node D, multiple CFOs will appear when
the oscillator of D is not perfectly matched to the oscillators of S and R or
there exists Doppler shift due to the mobility of D or R. Let $\varepsilon _1$
and $\varepsilon _2$ denote the frequency offsets of the direct link and relay link
respectively, which are normalized by the OFDM subcarrier spacing. Define
\[
c_i (\varepsilon _i ,n) \buildrel \Delta \over = \frac{1}{N}\exp ({j2\pi
\varepsilon _i n} \mathord{\left/ {\vphantom {{j2\pi \varepsilon _i n} N}}
\right. \kern-\nulldelimiterspace} N  ),\ \mbox{for} \
 i = 1,2.
\]
Then, the received signals of the direct link and relay link at destination node D can be
expressed as

\begin{equation}
\label{eq6}
y_{1 } [n] = Nc_1 (\varepsilon _1 ,n)(h_1 [n] \ast x[n]) + z_1 [n],
\end{equation}

\begin{equation}
\label{eq7}
y_{2 } [n] = \rho Nc_2 (\varepsilon _2 ,n)(h_2 [n]\ast h_3 [n]\ast x[n])
+ \rho z_{_2 } [n] + z_3 [n],
\end{equation}

\noindent
where $z_1 [n]$, $z_2 [n]$ and $z_3 [n]$ are AWGN with zero-mean and variance of $\sigma_ {Z_1}^2$, $\sigma_ {Z_2}^2$
and $\sigma_ {Z_3}^2$, $\ast$ denotes the convolution operator, and the convolution operation of
$x \left[ n \right]$ and $ y \left[ n \right]$ is defined as $
x \left[ n \right]\ast y \left[ n \right] = \sum\nolimits_{r = 0}^{N-1 }
{x [r] y[n - r]}$ .

\section{SNR Analysis}

   As we know, the CFO can be divided into an integer CFO (ICFO) and a fractional CFO (FCFO).
The ICFO results in the cyclic shift of the subcarriers, and
it should be corrected perfectly for the proper operation of the receiver. Thus, in
the following, we only consider the effect of the FCFO on the SNR performance in
OFDM based AF relay systems with the assumption that the ICFO has already been compensated.

    Applying discrete Fourier transform (DFT) to $y_{1 } [n]$ and $y_2
[n]$, we have

\begin{eqnarray}
Y_{1 }\left[ k \right] &=& C_1 \left( {\varepsilon _1 ,k} \right)\ast (H_1
\left[ k \right]X\left[ k \right]) + Z_{1 } \left[ k \right]
\nonumber \\
&=& C_1 \left( {\varepsilon _1 ,0} \right)H_1 \left[ k \right]X\left[ k
\right] + I_1 \left[ k \right] + Z_{1 }\left[ k \right],
\label{eq8}
\end{eqnarray}
\vspace*{-20pt}
\begin{eqnarray}
Y_2 [k] &=& \rho C_2 (\varepsilon _2 ,k)\ast
(H_2 [k]H_3 [k]X[k]) + \rho
Z_2 [k] + Z_3 [k]   \nonumber \\
&=& \rho C_2 \left( \varepsilon _2 ,0 \right)H_2 \left[ k \right]H_3 \left[ k
\right]X\left[ k \right] + \rho I_2 \left[ k \right] + \rho Z_2  \left[
k \right] \nonumber\\
&& + Z_3  \left[ k \right],
\label{eq9}
\end{eqnarray}

\noindent
where $H_i \left[ k \right]$ and $Z_i \left[ k \right]$ are the DFTs of
$h_i [n]$ and $z_i [n]$ for $i=1, 2, 3$, $C_i \left( {\varepsilon _i ,k} \right)$ is
the DFT of $c_i (\varepsilon _i ,n)$ which can be expressed as

\begin{eqnarray}
\lefteqn{
C_i (\varepsilon _i ,k) =  {\frac{\sin [\pi (\varepsilon _i -
k)]}{N\sin [{\pi (\varepsilon _i - k)} \mathord{\left/ {\vphantom {{\pi
(\varepsilon _i - k)} N}} \right. \kern-\nulldelimiterspace} N]}\exp \left[
{j\pi \left( {\varepsilon _i - k} \right)\left( {1 - \frac{1}{N} } \right)}
\right]} ,
}\quad\qquad\quad\qquad\quad\qquad\quad\qquad \nonumber\\
&& \mbox{for} \ i = 1, 2, \nonumber
\end{eqnarray}

\noindent
$I_1 \left[ k \right]$ and $I_2 \left[ k \right]$ are the inter-carrier interference
(ICI) which can be expressed as

\[
I_1 \left[ k \right] = \sum\limits_{r = 1}^{N - 1} {C_1 \left( {\varepsilon _1 ,r} \right)H_1 \left[ {k - r} \right]X\left[ {k - r} \right]},
\]

\[
I_2 \left[ k \right] = \sum\limits_{r = 1}^{N - 1} {C_2 \left( {\varepsilon _2 ,r} \right)
H_2 \left[ {k - r} \right]H_3 \left[ {k - r} \right]X\left[ {k - r} \right]}.
\]

    A coherent receiver should be able to estimate the phase of $C_1 \left(
{\varepsilon _1 ,0} \right)H_1 \left[ k \right]$ and $C_2 \left(
{\varepsilon _2 ,0} \right)H_2 \left[ k \right]H_3 \left[ k \right]$ in
order to decode the received signals correctly. Let

\[
C_1 \left( {\varepsilon _1 ,0} \right)H_1 \left[ k \right] = \vert C_1
\left( {\varepsilon _1 ,0} \right)H_1 \left[ k \right]\vert \exp [{j\beta _1
\left( k \right)}],
\]
\vspace*{-20pt}
\begin{eqnarray}
\lefteqn{
C_2 \left( {\varepsilon _2 ,0} \right)H_2 \left[ k \right]H_3 \left[ k
\right]
}\quad \nonumber \\
&& = \vert C_2 \left( {\varepsilon _2 ,0} \right)H_2 \left[ k
\right]H_3 \left[ k \right]\vert \exp [{j\beta _2 \left( k \right)}]. \nonumber
\end{eqnarray}

\noindent
Assume that $\beta _1 \left( k \right)$ and $\beta _2 \left( k \right)$
can be perfectly estimated, and define $R_1 \left[ k \right] = \exp [{
- j\beta _1 \left( k \right)}] Y_1 \left[ k \right]$, $R_2 \left[ k \right]
= \exp [{ - j\beta _2 \left( k \right)}] Y_2 \left[ k \right]$. Then, we have

\begin{equation}
\label{eq10}
R_1 \left[ k \right] = \vert C_1 \left( {\varepsilon _1 ,0} \right)H_1 \left[ k \right]\vert
X\left[ k \right] + I_1^{'} \left[ k \right] + Z_1^{'} \left[ k \right],
\end{equation}
\vspace*{-20pt}
\begin{eqnarray}
\label{eq11}
\lefteqn{
R_2 \left[ k \right] = \rho |C_2 \left( {\varepsilon _2 ,0} \right)H_2 \left[ k \right]H_3
\left[ k \right]|X\left[ k \right] + \rho I_2^{'} \left[ k \right]
} \quad  \nonumber \\
&& + \rho Z_2^{'} \left[ k \right] + Z_3^{'} \left[ k \right],
\end{eqnarray}

\noindent
where $Z_1^{'} \left[ k \right] = \exp \left[{ - j\beta _1 \left( k \right)} \right] Z_1 \left[ k
\right]$, $Z_2^{'} \left[ k \right] = \exp \left[ { - j\beta _2 \left( k \right)}\right]$$ Z_2 \left[ k
\right]$, $Z_3^{'} \left[ k \right] = \exp \left[ { - j\beta _2 \left( k \right)}\right] Z_3 \left[ k
\right]$, $I_1^{'} \left[ k \right] = \exp \left[{ - j\beta _1 \left( k \right)}\right] I_1 \left[ k
\right]$, $I_2^{'} \left[ k \right] = \exp \left[{ - j\beta _2 \left( k \right)}\right] I_2 \left[ k
\right]$. By employing equal gain combining (EGC), the decision metric $R[k]$ for coherent demodulation can
be expressed as
\begin{eqnarray}
R[k] &=& R_1 [k] + R_2 [k]
\nonumber \\
& = & \left( {\vert C_1 \left( {\varepsilon _1 ,0} \right)H_1 \left[
k \right]\vert + \rho \vert C_2 \left( {\varepsilon _2 ,0} \right)H_2 \left[
k \right]H_3 \left[ k \right]\vert } \right)X\left[ k \right] \nonumber\\
&& + I_1^{'} \left[ k \right] + \rho I_2^{'}
\left[ k \right]  + Z_1^{'} \left[ k \right] + \rho Z_2^{'} \left[ k \right] + Z_3^{'} \left[ k \right].
\end{eqnarray}

Define the SNR at destination node D as follows
\mathindent=0mm
\begin{eqnarray}
\label{eq13}
\lefteqn{
\mbox{SNR}
}\nonumber \\
 &&= {\mbox{E}\left\{ {\vert \left( {\vert C_1 \left( {\varepsilon _1 ,0}
\right)H_1 \left( k \right)\vert \!+ \! \rho \vert C_2 \left( {\varepsilon _2 ,0}
\right)H_2 \left( k \right)H_3 \left( k \right)\vert } \right)X\left( k
\right)\vert ^2} \right\}}\nonumber\\
&& /\  {\mbox{E}\left[ {\vert \left( {I_1^{'} \left[ k \right] + \rho I_2^{'} \left[ k \right]
+ Z_1 ^{'} \left[ k \right] \!+\!   \rho Z_2^{'} \left[ k
\right] + Z_3^{'}  \left[ k \right]} \right)\vert ^2} \right]}.
\end{eqnarray}
\mathindent=7mm

\noindent
Assume that all the considered channels are independent with each
other and their means are zero. Let

\[
\mbox{E}\left\{ {\vert X[k]\vert ^2} \right\} = \sigma _X^2 , \ \
\mbox{E}\left\{ {\vert H_i [k]\vert ^2} \right\} = \sigma _{ H_i }^2 , \
\mbox{for} \   i = 1,2,3.
\]

\noindent
Then, by exploiting the following relationship

\begin{equation}
\label{eq16}
\sum\limits_{r = 0}^{N - 1} {\vert C(\varepsilon _i ,r)\vert ^2} =
N\sum\limits_{r = 0}^{N - 1} {\vert c\left( {\varepsilon _i ,r} \right)\vert
^2} = 1, \ \mbox{for} \  i = 1,2,
\end{equation}

\noindent
the numerator and denominator of (\ref{eq13}) can be simplified as follows
\mathindent=0mm
\begin{eqnarray}
\lefteqn{
\mbox{E}\left\{ {\vert \left( {\vert C_1 \left( {\varepsilon _1 ,0} \right)H_1
\left( k \right)\vert + \rho \vert C_2 \left( {\varepsilon _2 ,0} \right)H_2
\left( k \right)H_3 \left( k \right)\vert } \right)X\left( k \right)\vert
^2} \right\}
}\quad \nonumber \\
&=&   \mbox{E}\left[ {\vert C_1 \left( {\varepsilon _1 ,0} \right)\vert ^2}
\right]\sigma _{H_1 }^2 \sigma _X^2 + \rho ^2\mbox{E}\left[ {\vert C_2 \left(
{\varepsilon _2 ,0} \right)\vert ^2} \right]\sigma _{H_2 }^2 \sigma _{H_3
}^2 \sigma _X^2 , \nonumber
\end{eqnarray}

\begin{eqnarray}
\lefteqn{
\mbox{E}\left[ {\vert \left( {I_1^{'} \left[ k \right] + \rho I_2^{'} \left[ k \right]
+ Z_1^{'} \left[ k \right] + \rho Z_2^{'} \left[ k \right] + Z_3^{'} \left[ k
\right]} \right)\vert ^2} \right]
}\quad \nonumber \\
&=&  \left\{ {1 - \mbox{E}\left[ {\vert C_1 \left( {\varepsilon _1 ,0} \right)\vert
^2} \right]} \right\}\sigma _{_{H_1 } }^2 \sigma _X^2
+ \rho ^2\left\{ {1 - \mbox{E}\left[ {\vert C_2 \left( {\varepsilon _2 ,0}
\right)\vert ^2} \right]} \right\}
\nonumber \\
&& \times \sigma _{_{H_2 } }^2 \sigma _{_{H_3 } }^2
\sigma _X^2 + \sigma _{Z_1 }^2 + \rho ^2\sigma _{Z_2 }^2 + \sigma _{Z_3 }^2. \nonumber
\end{eqnarray}
\mathindent=7mm

\noindent
Accordingly, the SNR in (\ref{eq13}) can be expressed as
\vspace*{-20pt}
\mathindent=0mm
%\begin{eqnarray}
%\lefteqn{
%\mbox{SNR}
%}\nonumber \\
%&& = \frac{\mbox{E}\left[ {\vert C_1 \left( {\varepsilon _1 ,0} \right)\vert ^2}
%\right]\sigma _{H_1 }^2 \sigma _X^2 + \rho ^2\mbox{E}\left[ {\vert C_2 \left(
%{\varepsilon _2 ,0} \right)\vert ^2} \right]\sigma _{H_2 }^2 \sigma _{H_3
%}^2 \sigma _X^2 }{\left\{ {1 - \mbox{E}\left[ {\vert C_1 \left( {\varepsilon _1 ,0}
%\right)\vert ^2} \right]} \right\}\sigma _{_{H_1 } }^2 \sigma _X^2 + \rho
%^2\left\{ {1 - \mbox{E}\left[ {\vert C_2 \left( {\varepsilon _2 ,0} \right)\vert
%^2} \right]} \right\}\sigma _{_{H_2 } }^2 \sigma _{_{H_3 } }^2 \sigma _X^2 +
%\sigma _{Z_1 }^2 + \rho ^2\sigma _{Z_2 }^2 + \sigma _{Z_3 }^2 }. \nonumber
%\end{eqnarray}

\begin{eqnarray}
\lefteqn{
\mbox{SNR}
}\nonumber \\
&& = {{\left\{\mbox{E}\left[ {\vert C_1 \left( {\varepsilon _1 ,0} \right)\vert ^2}
\right]\sigma _{H_1 }^2 \sigma _X^2 + \rho ^2\mbox{E}\left[ {\vert C_2 \left(
{\varepsilon _2 ,0} \right)\vert ^2} \right]\sigma _{H_2 }^2 \sigma _{H_3
}^2 \sigma _X^2 \right\}}}
\nonumber\\
&&  /\left\{ \left\{ {1 - \mbox{E}\left[ {\vert C_1 \left( {\varepsilon _1 ,0}
\right)\vert ^2} \right]} \right\}\sigma _{_{H_1 } }^2 \sigma _X^2 \right.  \nonumber\\
&&  + \rho
^2\left\{ {1 - \mbox{E}\left[ {\vert C_2 \left( {\varepsilon _2 ,0} \right)\vert
^2} \right]} \right\}\sigma _{_{H_2 } }^2 \sigma _{_{H_3 } }^2 \sigma _X^2 \nonumber\\
&& \left. + \sigma _{Z_1 }^2 + \rho ^2\sigma _{Z_2 }^2 + \sigma _{Z_3 }^2 \right\}. \nonumber
\end{eqnarray}
\mathindent=7mm

\noindent
For certain frequency offsets $\varepsilon _1 $ and $\varepsilon _2 $, the
\mbox{SNR} can be further simplified as follows,

\vspace*{-10pt}
\begin{eqnarray}
\label{eq26}
\lefteqn{
\mbox{SNR}\left( {\varepsilon _1 ,\varepsilon _2 } \right)
}\nonumber \\
&&
 = \left\{f_N^2 \left(
{\varepsilon _1 } \right)\sigma _{H_1 }^2 \sigma _X^2 + \rho ^2f_N^2 \left(
{\varepsilon _2 } \right)\sigma _{H_2 }^2 \sigma _{H_3 }^2 \sigma _X^2
\right\}\nonumber\\
&& / \left\{\left[ {1 - f_N^2 \left( {\varepsilon _1 } \right)} \right]\sigma _{_{H_1
} }^2 \sigma _X^2 + \rho ^2\left[ {1 - f_N^2 \left( {\varepsilon _2 }
\right)} \right]\sigma _{_{H_2 } }^2 \sigma _{_{H_3 } }^2 \sigma _X^2 \right. \nonumber\\
&&\left. + \sigma _{Z_1 }^2 + \rho ^2\sigma _{Z_2 }^2 + \sigma _{Z_3 }^2 \right\}
\end{eqnarray}

\noindent
where
$f_N \left( {\varepsilon _i } \right) = {\sin \left( {\pi \varepsilon
_i } \right)}/ \left[ N\sin \left( {\pi \varepsilon _i / N} \right)\right]
$ for $i = 1, 2$. From (\ref{eq26}), it is obvious that the existence of multiple CFOs
introduces SNR degradation, and it can be seen that the SNR decreases as the CFO $\varepsilon _1 $
or $\varepsilon _2 $ increases from 0 to 1/2 and that it is maximized if and
only if $\varepsilon _1 = \varepsilon _2 = 0$.

    Define the sensitivities of the SNR to $\varepsilon
_1 $ and $\varepsilon _2 $ as follows

\begin{equation}
\label{eq17}
\lambda _1 = \vert \frac{\partial \mbox{SNR}\left( {\varepsilon _1 ,\varepsilon _2
} \right)}{\partial \varepsilon _1 }\vert , \
\lambda _2 = \vert \frac{\partial \mbox{SNR}\left( {\varepsilon _1 ,\varepsilon _2
} \right)}{\partial \varepsilon _2 }\vert .
\end{equation}

\noindent
Then, we have

\begin{eqnarray}
\label{eq18}
\lambda _1 & =& \frac{2\vert f_N^{'} (\varepsilon _1 )\vert \sigma _{H_1 }^2
\sigma _X^2 (A + B)}{B^2}, \ \nonumber \\
\lambda _2 &=& \frac{2\vert f_N^{'} (\varepsilon _2 )\vert \rho ^2\sigma _{H_2
}^2 \sigma _{H_3 }^2 \sigma _X^2 (A + B)}{B^2},
\end{eqnarray}

\noindent
where A and B are the numerator and denominator of (\ref{eq26}), respectively.
By comparing $\lambda _1 $ with $\lambda _2 $ in (\ref{eq18}), it can be seen that the
sensitivities of the SNR to $\varepsilon _1 $ and $\varepsilon _2 $ are
different. The difference is mainly caused by the different power of the
corresponding link channel and gain factor.

    As far as the gain factor $\rho $ is concerned, it can be expressed as

\begin{equation}
\label{eq20}
\rho = \sqrt {\frac{P_{R} }{\sigma _{H_2 }^2 P_{S} + \sigma _{Z_2 }^2 }},
\end{equation}

\noindent
where $P_{R} $ and $P_{S} $ are the average transmitting power of R and
S, respectively. Under the power constraint $P = P_{S} + P_{R} $, many
power allocation (PA) schemes can be applied. Basically, they can be
classified as uniform power allocation (UPA) and optimal power allocation
(OPA). In the following, unless otherwise stated, UPA is adopted.
Nevertheless, it can be easily extended to OPA. Employing UPA, we obtain

\begin{equation}
\label{eq21}
\rho =
\sqrt {\frac{1}{\sigma _{H_2 }^2 + {2\sigma _{Z_2 }^2 } / {P}}} .
\end{equation}

\noindent
When $P \gg \sigma _{z_2 }^2 $, we have $\rho \approx {1 \mathord{\left/
 {\vphantom {1 {\sqrt {\sigma _{H_2 }^2 } } }} \right.
 \kern-\nulldelimiterspace} {\sqrt {\sigma _{H_2 }^2 } } }$.
By employing the above simplification, (\ref{eq26}) and (\ref{eq18}) can be
rewritten as follows

\begin{eqnarray}
\label{eq23}
\lefteqn{
\mbox{SNR}\left( {\varepsilon _1 ,\varepsilon _2 } \right)
} \nonumber\\
&& = \left\{f_N^2 \left(
{\varepsilon _1 } \right)\sigma _{H_1 }^2 \sigma _X^2 + f_N^2 \left(
{\varepsilon _2 } \right)\sigma _{H_3 }^2 \sigma _X^2 \right\} \nonumber\\
&& / \left\{
 \left[ {1 - f_N^2 \left( {\varepsilon _1 } \right)} \right]\sigma _{{H_1 }
}^2 \sigma _X^2 + \left[ {1 - f_N^2 \left( {\varepsilon _2 } \right)}
\right]\sigma _{{H_3 } }^2 \sigma _X^2 \right. \nonumber\\
&& \left. + \sigma _{Z_1 }^2 + \sigma _{H_2 }^2 \sigma _{Z_2 }^2 + \sigma
_{Z_3 }^2 \right\},
\end{eqnarray}

\vspace*{-20pt}
\begin{eqnarray}
\label{eq24}
\lambda _1 &= & \frac{2\vert f_N^{'} (\varepsilon _1 )\vert \sigma _{H_1 }^2
\sigma _X^2 (A + B)}{B^2},
\nonumber \\
\lambda _2 &=& \frac{2\vert f_N^{'} (\varepsilon _2 )\vert \sigma _{H_3 }^2
\sigma _X^2 (A + B)}{B^2}.
\end{eqnarray}

\noindent
From (\ref{eq24}), it can be seen that the sensitivity of the SNR to
$\varepsilon _1 $ and $\varepsilon _2 $ is determined only by the power of
the direct link channel and that of the second hop of the relay link channel
when UPA is employed.

\begin{figure}[!t]
\centerline{\includegraphics[scale=0.8]{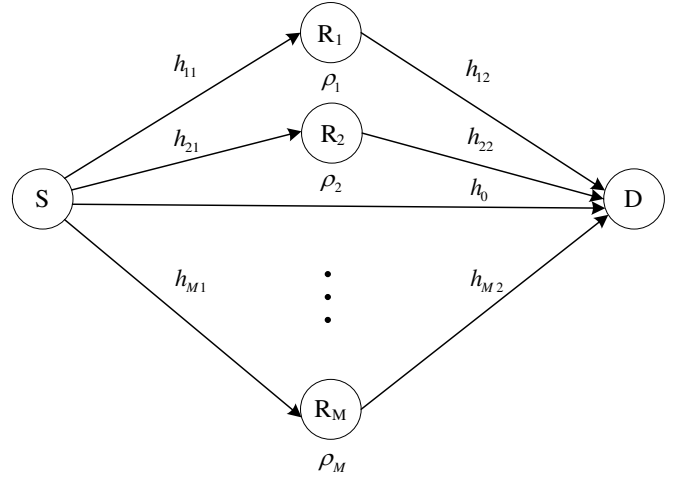}}
\caption{The Schematic of a multiple-relay-node AF system.}
\label{fig2}
\end{figure}

    Furthermore, according to the above analysis, we can
easily extend the SNR expression in (\ref{eq26}) to the multiple-relay-node
scenario as depicted in Fig. \ref{fig2} as follows

\begin{eqnarray}
\label{eq27}
\lefteqn{
\mbox{SNR}\left( {\varepsilon _0 ,\varepsilon _1 ,...,\varepsilon _M } \right)
}\nonumber\\
&& =
\left\{f_N^2 \left( {\varepsilon _0 } \right)\sigma _{H_0 }^2 \sigma _X^2 +
\sum\limits_{i = 1}^M {\rho _i^2 f_N^2 \left( {\varepsilon _i }
\right)\sigma _{H_{i1} }^2 \sigma _{H_{i2} }^2 } \sigma _X^2 \right\}\nonumber\\
&& / \left\{\left[ {1 - f_N^2 \left( {\varepsilon _0 } \right)} \right]\sigma _{_{H_0 }
}^2 \sigma _X^2 + \sum\limits_{i = 1}^M {\rho _i^2 \left[ {1 - f_N^2 \left(
{\varepsilon _i } \right)} \right]\sigma _{_{H_{i1} } }^2 \sigma _{_{H_{i2}
} }^2 } \sigma _X^2 \right. \nonumber\\
&& \left. + \sigma _{Z_0 }^2 + \sum\limits_{i = 1}^M {\left(
{\sigma _{Z_{i1} }^2 + \rho _i^2 \sigma _{Z_{i2} }^2 } \right)} \right\}.
\end{eqnarray}

\section{Simulation Results}

In this section, simulation results are given to verify the analytical results
under the following conditions: $\sigma _{z_1 }^2 =
\sigma _{z_2 }^2 = \sigma _{z_3 }^2 $ and $\sigma _{H_3 }^2 = 4 \sigma _{H_1
}^2 $.

\begin{figure}[!t]
\centerline{\includegraphics[scale=0.85]{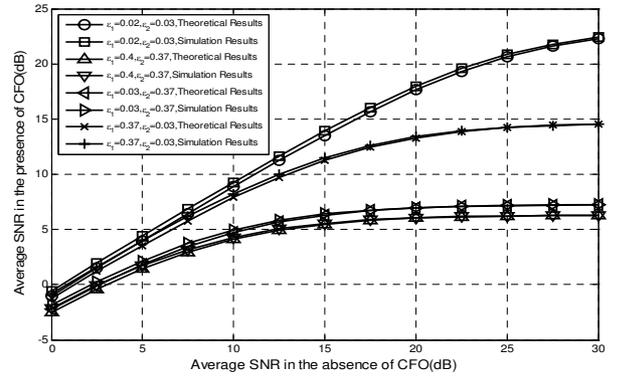}}
\caption{Average SNR in the presence of CFO for flat
fading channels.}
\label{fig3}
\end{figure}

\begin{figure}[!t]
\centerline{\includegraphics[scale=0.75]{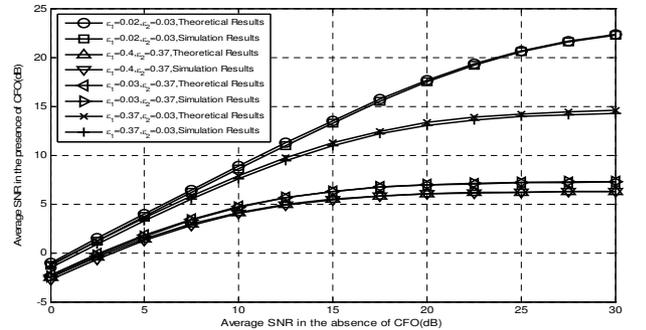}}
\caption{Average SNR in the presence of CFO for
frequency-selective fading channels.}
\label{fig4}
\end{figure}

Fig. \ref{fig3} presents the average SNR in the presence of CFO
for flat fading channels, while Fig. \ref{fig4} shows the average SNR in the presence
of CFO for frequency-selective fading channels. It can be seen
that the simulation results and theoretical results match very well. When
the CFOs increase, the SNR decreases dramatically. The SNR is more sensitive
to $\varepsilon _2 $, and the SNR degradation due to the CFOs is more serious in the high SNR
region. Compared with the flat fading channel, the
frequency-selective fading channel shows slightly larger SNR degradation.

\section{Conclusions}

In this letter, we have analyzed the effect of the CFOs on OFDM based AF relay systems.
From the derived SNR expression, it has been found that the SNR decreases
monotonically as the frequency offsets increase. Both analytical and numerical results
have shown that the SNR degradation due to the CFOs at high SNR values is larger than that at low SNR values
and that the sensitivity of the SNR to CFO is mainly determined
by the power of the corresponding link channel and gain factor.

\section*{Acknowledgments}
The work of Yanxiang Jiang, Yanxing Hu, and Xiaohu You was supported
in part by the Key National Project (No. 2009ZX03003-004-02), in
part by National Natural Science Foundation of China (No. 60702028),
in part by the Research Fund of Southeast University (No.
9204000033), and in part by the Research Fund of National Mobile
Communications Research Laboratory, Southeast University (No.
2009B03).

%\bibliographystyle{ieicetr}% bib style
%\bibliography{}% your bib database

%\begin{thebibliography}{99}% more than 9 --> 99 / less than 10 --> 9
%\bibitem{}
%\end{thebibliography}

%\profile{}{}
%\profile*{}{}% without picture of author's face

\end{document}